\documentclass[%
aps,%
prd,%
reprint,%
showpacs,%
amsmath,%
amssymb,%
nofootinbib%
]{revtex4-1}

\usepackage{amsfonts}
\usepackage{graphicx}
\usepackage{xcolor}
\usepackage{mathtools}
\usepackage{bm}
\usepackage{dsfont}
\usepackage{hyperref}

\newcommand{\be}{\begin{equation}}
\newcommand{\ee}{\end{equation}}
\newcommand{\bg}{\bar{g}}
\newcommand{\mn}{{\mu\nu}}
\newcommand{\rs}{{\rho\sigma}}
\newcommand{\diag}{\text{diag}}

\newcommand{\On}{\text{O}(n)}
\newcommand{\ccr}{c_\text{crit}}
\DeclareMathOperator{\Tr}{Tr}
\newcommand{\sg}{\sqrt{g}}
\newcommand{\sbg}{\sqrt{\bg}}
\newcommand{\sbgx}{\sqrt{\bg(x)}}
\newcommand{\sbgy}{\sqrt{\bg(y)}}
\newcommand{\Rk}{\mathcal{R}_k}
\newcommand{\td}{\text{d}}
\newcommand{\dd}{\td^d}
\newcommand{\mO}{\mathcal{O}}
\newcommand{\bxi}{\bar{\xi}}
\newcommand{\half}{{\textstyle\frac{1}{2}}}

\begin{document}

%-------------------------------------------------------------------------------------------------------
\title{Field Parametrization Dependence in Asymptotically Safe Quantum Gravity}

\author{Andreas Nink}
\email[]{nink@thep.physik.uni-mainz.de}
\affiliation{PRISMA Cluster of Excellence, Institute of Physics, Johannes Gutenberg University Mainz,
Staudingerweg 7, 55099 Mainz, Germany}

\preprint{MITP/14-086}
%-------------------------------------------------------------------------------------------------------

\begin{abstract}
Motivated by conformal field theory studies we investigate Quantum Einstein Gravity with
a new field parametrization where the dynamical metric is basically given by the exponential
of a matrix-valued fluctuating field, $g_{\mu\nu}=\bg_{\mu\rho}(e^h)^\rho{}_\nu$.
In this way, we aim to reproduce the critical value of the central charge when considering
$2+\epsilon$ dimensional spacetimes. With regard to the Asymptotic Safety program,
we take special care of possible fixed points and new structures of the corresponding RG
flow in $d=4$ for both single- and bi-metric truncations. Finally, we discuss the issue of
restoring background independence in the bi-metric setting.

\end{abstract}

\maketitle

%-------------------------------------------------------------------------------------------------------

\section{Introduction}

Conventional quantum field theories in flat space require the Minkowski metric $\eta_\mn$ as
an indispensable background structure in order to introduce a notion of time and causality,
and to construct actions with interesting ``nontopological'' and covariant terms. Many concepts
known for the flat case can be generalized and transferred to curved spacetimes, where the metric
$g_\mn$ assumes the role of the crucial background arena all invariants of the theory can be
constructed in. In quantum gravity, however, there is a fundamental conceptual difficulty since
the metric itself is a dynamical field now, having the consequence that, a priori, the arena does
not exist.

An elegant way out of this problem is provided by the introduction of a nondynamical
background metric $\bg_\mn$ that is kept arbitrary and that serves as the basis of nontopological
covariant constructions. The dynamical metric $g_\mn$ is then parametrized by a combination of this
background and some dynamical field(s). The crucial idea is that -- due to the arbitrariness of
the background -- in the end all physical quantities like scattering amplitudes must not depend on
$\bg_\mn$ any more. This bootstrap argument is one implementation of \emph{background independence},
a property that must be satisfied by any meaningful theory of quantum gravity. Its application has
led to great successes in many different physical situations.

As a consequence of background arbitrariness, also the way in which the dynamical metric is
parametrized is not determined without assuming or knowing the fundamental degrees of
freedom and without having defined a fluctuating field. The most famous choice of parametrization
is the standard background field method, where the dynamical field undergoes a \emph{linear split}
into background plus fluctuation \cite{DW03}. This has turned out to be a powerful technique in
many quantum field theory calculations, in particular in non-Abelian gauge theories. In the context
of gravity it reads
\be
g_\mn = \bg_\mn + h_\mn \, ,
\label{eq:stdParam}
\ee
where the fluctuating field $h_\mn$ is a symmetric tensor. In the following we refer to this
split as \emph{standard parametrization}.

As an example for a different choice of parametrization we can consider the nonlinear relation
\cite{FP94}
\be
g_\mn = e^a{}_\mu e^b{}_\nu\, \bg_{ab} \, ,
\ee
with invertible matrix-valued fields $e^a{}_\mu$. In the vielbein formalism for instance
\cite{HR12,DP13}, the dynamical metric $g_\mn$ is parametrized by $g_\mn = e^a{}_\mu e^b{}_\nu\,
\eta_{ab}$, together with $e^a{}_\mu = \bar{e}^a{}_\mu + \varepsilon^a{}_\mu$, so the fluctuations
$\varepsilon^a{}_\mu$ around the background field $\bar{e}^a{}_\mu$ contribute nonlinearly to
$g_\mn$.

Recently yet another parametrization has attracted increasing interest \cite{KKN93}. Although it
is a nonlinear relation, too, it does not introduce more independent components than contained in
$g_\mn$. Here, the metric is determined by the \emph{exponential} of a fluctuating field,
\be
g_{\mu\nu}=\bg_{\mu\rho}(e^h)^\rho{}_\nu \, .
\label{eq:newParam}
\ee
Again, $\bg_{\mu\rho}$ denotes the background metric, and $h$ is a symmetric matrix-valued field,
$h_\mn = h_{\nu\mu}$ (or $h^\mu{}_\nu = h_\nu{}^\mu$ with the shifted index position). As usual,
indices are raised and lowered by means of the background metric. In this work we refer to the
exponential relation \eqref{eq:newParam} as \emph{new parametrization}.

A priori, there seems to be no reason to prefer one parametrization over another one. It is well
known that field redefinitions in the path integral for the partition function do not change
S-matrix elements \cite{CWZ69}. While this equivalence theorem is based on the use of the equations
of motion, the (off shell) effective action $\Gamma$ in the usual formulation does still depend on
the choice of the parametrization.

In a geometric setting, by regarding the configuration space as a manifold, Vilkovisky and DeWitt
constructed an effective action in a covariant way such that it is independent of parametrizations
and gauge conditions for the quantized fields both off and on shell \cite{V84}. In this approach,
however, $\Gamma$ can have a remaining dependence on the chosen configuration space metric
\cite{O91}. Furthermore, unlike the conventional effective action, the Vilkovisky-DeWitt effective
action does not generate the 1PI correlation functions; instead, it is governed by modified Ward
identities \cite{BK87} which, in the present context, would relate $\delta\Gamma/\delta g_\mn$ to
$\delta\Gamma/\delta \bg_\mn$. This is a first hint that specific parametrizations can
be of interest as they have an effect on off shell quantities, and appropriate choices may
simplify calculations. An example is the frame dependence in cosmology which has been investigated
in reference \cite{KS14} at one-loop level.

Studies (without the Vilkovisky-DeWitt approach) of the renormalization group (RG) show that
$\beta$-functions and fixed points can vary when the parametrization is changed \cite{W74,M98}.
In addition, parametrization invariance is violated even on shell when truncations, e.g.\
derivative expansions, are considered \cite{M98}. Combining RG techniques with the ideas of
Vilkovisky and DeWitt leads to the geometrical effective average action, which is constrained
by generalized modified Ward identities \cite{P03}. Therefore, again, parametrization and gauge
invariance can be obtained only at the expense of nontrivial dependencies on the background.
In summary, off shell quantities in both conventional and Vilkovisky-DeWitt approach can depend
on the underlying parametrization and/or the background.

Thus, it is usually safer to consider physical observables as they should not exhibit any
parametrization or gauge dependence. In quantum gravity, however, it is not even clear what
physically meaningful observable quantities are, and so far there is no experiment for a direct
measurement of quantum gravity effects \cite{W09}. Based on effective field theory arguments
it is possible to compute the leading quantum corrections to the Newtonian potential \cite{D94},
but the effect is unobservably small and the description is valid only in the low energy regime,
so it cannot be considered a fundamental theory of the gravitational field. Due to the problem
of finding observable quantities, the best one can do with a candidate theory of quantum gravity
is to test it for self-consistency, check the classical limit, and compare it with other
approaches. In this regard it is of substantial interest to study off shell quantities like
$\beta$-functions. Their parametrization dependence might then be exploited to simplify the
comparison between different theories. For instance, if correlation functions of vertex
operators of the type $e^{ikX}$ in string theory \cite{GSW87} are supposed to be compared
with another approach, it may be natural to use the exponentials of some fields there as well.

In the present work we use a fully nonperturbative framework to show within the
Einstein--Hilbert truncation that $\beta$-functions do indeed depend on the parametrization,
and the exponential relation \eqref{eq:newParam} turns out to be more appropriate for a
comparison with conformal field theories. Particular attention is paid to RG fixed points
in the context of Asymptotic Safety. It is assumed that the reader is familiar with the
concept; for introductions and reviews of Asymptotic Safety see \cite{ASReviews}.
At this point we want to make an important remark. Apart from the fact that a
reparametrization can change objects whose direct physical meaning is obscured, it could
also give rise to a fundamental change: In principle, there is the possibility that in
parametrization A the defining path integral has a suitable continuum limit according to
the Asymptotic Safety scenario, i.e.\ coupling constants approach a fixed point in the UV,
while there may not be such a well defined limit for parametrization B. However, when
resorting to truncations, it would be hard to decide whether such a change due to
reparametrization is actually fundamental or just a truncation artifact. One could find
that a suitable fixed point is absent in one parametrization, while it exists in another
one, but after enlarging the truncated theory space the resulting differences between the
two parametrizations might diminish eventually. Clearly, in that case higher order truncations
would have to be considered to obtain more reliable results.

In the following, we focus more concretely on the properties of the ``new'' exponential
parametrization \eqref{eq:newParam}. There are several independent reasons that strongly motivate
its use.

\textbf{i)} The first argument is a geometric one. We answer the question if the right hand side
of relation \eqref{eq:newParam} represents a metric. As we prove in appendix \ref{app:Logs}, there
is a \emph{one-to-one correspondence} between dynamical metrics $g_\mn$ and symmetric matrices $h$.
(Note that we consider Euclidean signature spacetimes throughout this paper, so metrics are positive
definite.) That means that, given a dynamical and a background metric, $g_\mn$ and $\bg_\mn$,
respectively, there exists a unique symmetric matrix $h$ satisfying equation \eqref{eq:newParam}.
If, on the other hand, $\bg_\mn$ and a symmetric $h_\mn$ are given, then $g_\mn$ defined by
$g_\mn=\bg_{\mu\rho}(e^h)^\rho{}_\nu$ is symmetric and positive definite, so it is again an admissible
metric. As a consequence, a path integral over $h_\mn$ captures all possible $g_\mn$, and no $g_\mn$
is counted twice or even more times. The matrices $h$ can be seen as tangent vectors corresponding
to the space of metrics and equation \eqref{eq:newParam} as the exponential map (even though not
using the Vilkovisky-DeWitt connection). Due to the positive definiteness of $g_\mn$ guaranteed by
construction, the new parametrization seems to be preferable to the standard one given by
equation \eqref{eq:stdParam}.

\textbf{ii)} Our main motivation comes from an apparent connection to conformal field theories (CFT).
To see this, we examine a path integral for two dimensional gravity coupled to conformal matter
(i.e.\ to a matter theory that is conformally invariant when the metric is flat) with central
charge $c$. Here it is sufficient to consider matter actions constructed from scalar fields. Then
$c$ is just the number of these scalar fields. As shown by Polyakov \cite{P81}, the path integral
decomposes into a path integral over the conformal mode $\phi$ with a Liouville-type action times a
$\phi$-independent part. Owing to the integral over Faddeev-Popov ghosts, the kinetic term for $\phi$
comes with a factor of $(c-26)$, reflecting the famous critical dimension of string theory. If,
finally, the implicit $\phi$-dependence of the path integral measure is shifted into the action, the
kinetic term for $\phi$ gets proportional to $(c-25)$ \cite{DDK88}. For this reason we call
\be
\ccr=25 
\ee
the \emph{critical central charge} at which $\phi$ decouples.

How is that related to Asymptotic Safety? Let us consider the RG running of the dimensionless
version of Newton's constant, $g$, now slightly away from two dimensions, $d=2+\epsilon$. Already
a perturbative treatment shows that the $\beta$-function has the general form
\be
 \beta_g = \epsilon g - b g^2 ,
\label{eq:betaeps}
\ee
up to order $\mathcal{O}(g^3)$ \cite{W80}, leading to the non-Gaussian fixed point
\be
 g_* = \epsilon/b .
\ee
It turns out that the coefficient $b$ depends on the underlying parametrization of the metric.
Perturbative calculations based on the standard parametrization \eqref{eq:stdParam} yield
$b=\frac{2}{3}(19-c)$, where $c$ denotes again the number of scalar fields \cite{W80,T77}.
This gives rise to the critical central charge
\be
 \ccr=19
\label{eq:ccritStdPert}
\ee
in the standard parametrization. If, on the other hand, the new parametrization \eqref{eq:newParam}
underlies the computation, the critical central charge amounts to \cite{KKN93}
\be
 \ccr=25.
\label{eq:ccritNewPert}
\ee
Since many independent derivations yield $\ccr=25$, too, it appears ``correct'' in a certain sense.
This result seems to be another advantage of the exponential parametrization. In the present
work we investigate if it can be reproduced in a nonperturbative setup.

\textbf{iii)} Let us come back to an arbitrary dimension $d$. Parametrizing the metric with an
exponential allows for an easy treatment of the conformal mode which can be separated as the
trace part of $h$ in equation \eqref{eq:newParam}: We split $h_\mn$ into trace and traceless
part, $h_\mn=\hat{h}_\mn+\frac{1}{d}\bg_\mn \phi$, where $\phi=\bg^\mn h_\mn$ and
$\bg^\mn\hat{h}_\mn=0$. In this case, equation \eqref{eq:newParam} becomes
\be
g_{\mu\nu}=\bg_{\mu\rho}\big(e^{\hat{h}}\big)^\rho{}_\nu \, e^{\frac{1}{d}\phi} ,
\ee
so the trace part of $h$ gives a conformal factor. Using the matrix relation $\det(\exp M)=\exp(\Tr M)$
we obtain
\be
\sg = \sbg \, e^{\frac{1}{2}\phi} \, ,
\label{eq:sqrtg}
\ee
where $g$ ($\bg$) denotes the determinant of $g_\mn$ ($\bg_\mn$). The traceless part of $h$ has
completely dropped out of equation \eqref{eq:sqrtg}. Hence, unlike for the standard parametrization,
the cosmological constant appears as a coupling in the conformal mode sector only. This will become
explicit in our calculations.

\textbf{iv)} The new parametrization might simplify computations and cure singularities that are
possibly encountered with the standard parametrization. Here we briefly mention three examples.
(a) The RG flow of nonlocal form factors appearing in a curvature expansion of the effective
average action in $2+\epsilon$ dimensions is divergent in the limit $\epsilon\rightarrow 0$ for small
$k$ when based on \eqref{eq:stdParam} but has a meaningful limit with equation \eqref{eq:newParam}
\cite{SCM10}. (b) Similarly, when trying to solve the flow equations in scalar-tensor theories
\cite{NP09} in $d=3$ and $d=4$, singularities occurring in a standard calculation can be avoided by
using the new parametrization.\footnote{The author would like to thank R.~Percacci for pointing
this out.} (c) Related to argument iii) the exponential parametrization provides an easy access
to unimodular gravity \cite{E13}.

In this work we present a nonperturbative derivation of the $\beta$-functions of Newton's constant
and the cosmological constant. For that purpose, we study the effective average action within the
Einstein--Hilbert truncation (without using the Vilkovisky-DeWitt method), where the metric is
replaced according to equation \eqref{eq:newParam}. We will show that the results for the exponential
parametrization are significantly different from the ones obtained with the standard split
\eqref{eq:stdParam}. Although we encounter a stronger scheme dependence that has to be handled with
care, the favorable properties of the new parametrization seem to prevail, as discussed in the
final section.

%-------------------------------------------------------------------------------------------------------

\section{Framework}
\label{sec:framework}

We employ functional RG techniques to evaluate $\beta$-functions in a nonperturbative way. The method
is based upon the effective average action $\Gamma_k$, a scale dependent version of the usual
effective action $\Gamma$. By definition, its underlying path integral contains a mass-like regulator
function $\Rk(p^2)$ such that quantum fluctuations with momenta below the infrared cutoff scale $k$
are suppressed while only the modes with $p^2>k^2$ are integrated out. Thus, $\Gamma_k$ interpolates
between the microscopic action at $k\rightarrow\infty$ and $\Gamma$ at $k=0$. Its scale dependence
is governed by an exact functional RG equation (FRGE) \cite{W93,R98,M94},
\be
k \partial_k \Gamma_k = \frac{1}{2}\, \Tr \bigg[ \Big(\Gamma_k^{(2)}+ \Rk\Big)^{-1}\, 
k \partial_k \Rk\,\bigg] ,
\label{eq:FRGE}
\ee
where $\Gamma_k^{(2)}$ denotes the second functional derivative with respect to the fluctuating
field ($h_\mn$ in our case). In the terminology of reference \cite{CPR09} we choose
a type Ia cutoff, i.e.\ $\Rk$ is a function of the covariant Laplacian.

As outlined above, any field theoretic description of quantum gravity requires the introduction of
a background metric. Consequently, $\Gamma_k$ is a functional of both $g_\mn$ and $\bg_\mn$ in
general, i.e.\ $\Gamma_k\equiv\Gamma_k[g,\bg]$. In terms of $h$ the two parametrizations give rise
to the functionals
\be
\Gamma_k^\text{standard}[h;\bg] \equiv \Gamma_k[\bg+h,\bg] ,
\label{eq:GammaStandard}
\ee
as opposed to
\be
\Gamma_k^\text{new}[h;\bg] \equiv \Gamma_k\big[\bg \, e^h,\bg\big] .
\label{eq:GammaNew}
\ee
(We adopt the comma notation for $\Gamma_k[g,\bg]$ and the semicolon notation for
$\Gamma_k[h;\bg]$.)
The difference between \eqref{eq:GammaStandard} and \eqref{eq:GammaNew} is crucial: Since
the second derivative $\Gamma_k^{(2)}$ in \eqref{eq:FRGE} is with respect to $h$, the two
parametrizations give rise to different terms according to the chain rule,
\be
\begin{split}
 &\Gamma_k^{(2)}(x,y) \equiv \frac{1}{\sbgx\sbgy}\;\frac{\delta^2\Gamma_k}{\delta h(x) \,
      \delta h(y)} \\
   &= \frac{1}{\sbgx\sbgy}\int_u\int_v\, \frac{\delta^2\Gamma_k}{\delta g(u)\,\delta g(v)} \,
      \frac{\delta g(v)}{\delta h(x)}\,\frac{\delta g(u)}{\delta h(y)} \\
   &  \qquad +\frac{1}{\sbgx\sbgy}\int_u\, \frac{\delta \Gamma_k}{\delta g(u)} \,
      \frac{\delta^2 g(u)}{\delta h(x)\,\delta h(y)} \; ,
\end{split}
\label{eq:2ndVar}
\ee
where we suppressed all spacetime indices and used the shorthand $\int_u \equiv \int \dd u$.
The first term on the right hand side of equation \eqref{eq:2ndVar} is the same for both
parametrizations, at least at lowest order, because $\delta g_\mn(x)/\delta h_{\rho\sigma}(y)
= \delta^\rho_\mu \, \delta^\sigma_\nu \, \delta(x-y)$ in the standard case, and
$\delta g_\mn(x)/\delta h_{\rho\sigma}(y)\allowbreak = \delta^\rho_\mu \, \delta^\sigma_\nu \,
\delta(x-y) +\mO(h)$ with the new parametrization. The last term in \eqref{eq:2ndVar}, however,
vanishes identically for parametrization \eqref{eq:stdParam} since $\delta^2 g/\delta h^2 = 0$,
whereas the exponential relation \eqref{eq:newParam} entails
\begin{align}
&\frac{\delta^2 g_\mn(u)}{\delta h_{\rho\sigma}(x) \, \delta h_{\lambda\gamma}(y)} \\ 
&= {\textstyle \frac{1}{2}}
\left(\bg^{\sigma\lambda} \delta^\rho_\mu \, \delta^\gamma_\nu + \bg^{\rho\gamma} \delta^\lambda_\mu
\, \delta^\sigma_\nu \right) \delta(u-x) \delta(u-y) + \mO(h), \nonumber
\end{align}
leading to additional contributions to the FRGE \eqref{eq:FRGE}. Note that these new contributions
are proportional to the first variation of $\Gamma_k$ in \eqref{eq:2ndVar}, so the exponential
parametrization gives the same result as the standard one when going on shell. But, due to the
inherent off shell character of the FRGE, we expect differences in $\beta$-functions and the
corresponding RG flow.

Finally, let us comment on gauge invariance and fixing. Starting from relation \eqref{eq:newParam},
we observe that $(e^h)^\rho{}_\nu$ must transform as a tensor under general coordinate
transformations, if $g_\mn$ and $\bg_{\mu\rho}$ transform as tensors. It is possible to show then
that $h_\mn$ transforms in the same way, i.e.\
\be
\delta \mkern1mu h_\mn = \mathcal{L}_\xi h_\mn
\ee
under diffeomorphisms generated by the vector field $\xi$ via the Lie derivative $\mathcal{L}_\xi$.
In order to be as close to the standard calculations based on \eqref{eq:stdParam} as possible
\cite{R98}, we shall employ an analogous gauge fixing procedure. This can most easily be done by
observing that $h_\mn$ in the standard gauge fixing condition $\mathcal{F}_\alpha^\mn[\bg]
\, h_\mn=0$ can be replaced by $g_\mn$: We use the most convenient class of $\mathcal{F}$'s
where $\mathcal{F}_\alpha^\mn[\bg]$ is proportional to the covariant derivative $\bar{D}_\mu$
corresponding to the background metric, and therefore, $\mathcal{F}_\alpha^\mn[\bg] \, g_\mn
= \mathcal{F}_\alpha^\mn[\bg](\bg_\mn+h_\mn) = \mathcal{F}_\alpha^\mn[\bg] \, h_\mn = 0$ for
the standard parametrization.

Passing on to the new parametrization, we can choose the $g_\mn$-version of the gauge condition,
too, $\mathcal{F}_\alpha^\mn[\bg] \, g_\mn = 0$. This version is preferred to the one acting on
$h_\mn$ because, (a) it is hard to solve the true or ``quantum'' gauge transformation law for
$\delta h_\mn$ (by solving $\delta g_\mn = \mathcal{L}_\xi g_\mn$ while $\delta \bg_\mn=0$),
and (b) the $g_\mn$-choice leads to the same Faddeev-Popov operator as in the standard case
\cite{R98}. As a consequence, all contributions to the FRGE coming from gauge fixing and ghost
terms are the same for both parametrizations. By virtue of the one-to-one correspondence between
$g_\mn$ and $h_\mn$ (see appendix \ref{app:Logs}) this gauge fixing method is perfectly admissible.

We present a single-metric computation in section \ref{sec:single} and a bi-metric
\cite{bimetric,BR14} analysis in section \ref{sec:bi}. In the single-metric case, we employ
the harmonic gauge condition, $\mathcal{F}_\alpha^\mn[\bg] \, g_\mn = 0$ with
$\mathcal{F}_\alpha^\mn[\bg] =\delta^\nu_\alpha \,\bg^{\mu\rho}\bar{D}_\rho - \frac{1}{2} \,
\bg^\mn \bar{D}_\alpha$ (corresponding to $\rho=\frac{d}{2}-1$ in \cite{CPR09}), together with
a Feynman-type gauge parameter, $\alpha=1$. The bi-metric results are obtained by using the
$\Omega$ deformed $\alpha=1$ gauge \cite{BR14}. To summarize, we repeat the calculations of
\cite{R98} and \cite{BR14} with the new parametrization, where the modifications originate
from the gravitational part of $\Gamma_k$, while gauge fixing, ghost and cutoff contributions
remain the same.

%-------------------------------------------------------------------------------------------------------

\section{Results: Single-metric}
\label{sec:single}

As usual, we resort to evaluations of the RG flow within subspaces of reduced dimensionality, i.e.\
we truncate the full theory space. Our single-metric results are based on the Einstein--Hilbert
truncation \cite{R98},
\be
\begin{split}
\Gamma_k\big[g,\bg,\xi,\bxi\, \big] = \;&\frac{1}{16\pi G_k} \int \! \dd x \sg \,
	\big( -R + 2\Lambda_k \big) \\
	& + \Gamma_k^\text{gf}\big[g,\bg\big]
	+\Gamma_k^\text{gh}\big[g,\bg,\xi,\bxi\, \big].
\end{split}
\label{eq:EHtrunc}
\ee
Here $G_k$ and $\Lambda_k$ are the dimensionful Newton constant and cosmological constant,
respectively, $\Gamma_k^\text{gf}=\frac{1}{2\alpha}\frac{1}{16\pi G_k}\int\dd x \sbg \,
\bg^{\alpha\beta}\big(\mathcal{F}_\alpha^\mn g_\mn\big)\big(\mathcal{F}_\beta^{\rho\sigma}
g_{\rho\sigma}\big)$ is the gauge fixing action with $\mathcal{F}_\alpha^\mn$ and $\alpha$ as
given in the previous section, and $\Gamma_k^\text{gh}$ denotes the corresponding ghost action
with ghost fields $\xi$ and $\bxi$. After having inserted the respective metric parametrization
into the ansatz \eqref{eq:EHtrunc}, the FRGE \eqref{eq:FRGE} can be used to extract
$\beta$-functions.

%-------------------------------------------------------------------------------------------------------

\subsection{Known results for the standard parametrization}

For comparison, we begin by quoting known results for the standard parametrization.

\begin{figure}[tp]
  \centering
  \includegraphics[width=.78\columnwidth]{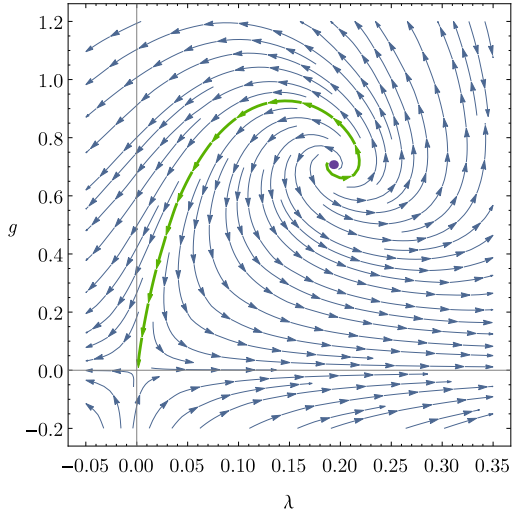}
  \caption{Flow diagram for the Einstein--Hilbert truncation in $d=4$ based on the \emph{standard
	  parametrization}. (First obtained in \cite{RS02} for
	  a sharp cutoff, here for the optimized cutoff \cite{L01}.)}
  \label{fig:StdSingle}
\end{figure}

In $4$ dimensions the resulting $\beta$-functions for the dimensionless couplings $g_k=k^{d-2}G_k$
and $\lambda_k=k^{-2}\Lambda_k$ \cite{R98} give rise to the flow diagram shown in figure
\ref{fig:StdSingle}. In addition to the Gaussian fixed point at the origin there exists a
non-Gaussian fixed point (NGFP) with positive Newton constant, suitable for the Asymptotic Safety
scenario \cite{W80}. It is also crucial that there are trajectories emanating from the NGFP and
passing the classical regime close to the Gaussian fixed point \cite{RW04}. (In figure
\ref{fig:StdSingle} one can see the separatrix, a trajectory connecting the non-Gaussian to the
Gaussian fixed point.) It has turned out that the qualitative picture (existence of NGFP, number
of relevant directions, connection to classical regime) is extremely stable under many kinds of
modifications of the setup (truncation ansatz, gauge, cutoff, inclusion of matter, etc.);
for reviews see \cite{ASReviews}. In particular, changes in the cutoff shape function do not
alter the picture, except for insignificantly shifting numerical values like fixed point
coordinates.

In $d=2+\epsilon$ dimensions the $\beta$-function of $g_k$ has the same structure as in the
perturbative analysis, see equation \eqref{eq:betaeps}, $\beta_g = \epsilon g - b g^2$. It is
possible to show that the coefficient $b$ is a universal number, i.e.\ it is independent of the
shape function, and its value is given by $b=\frac{38}{3}$ \cite{R98}. If, additionally, scalar
fields are included, then it reads $b=\frac{2}{3}(19-c)$, where $c$ denotes the number of scalar
fields. Thus, the standard parametrization gives rise to the \emph{universal} number for the
\emph{critical central charge}
\be
\ccr=19 ,
\ee
in agreement with the perturbative result \eqref{eq:ccritStdPert}.

%-------------------------------------------------------------------------------------------------------

\subsection{Results for the new parametrization}

We refrain from presenting details of the calculation and specify some intermediate results and
$\beta$-functions in appendix \ref{app:Details} instead.

%-------------------------------------------------------------------------------------------------------

\subsubsection*{Results in \texorpdfstring{$2+\epsilon$}{2+epsilon} dimensions}

Considering equations \eqref{eq:beta_g_FRG} and \eqref{eq:beta_lambda_FRG} for $d=2+\epsilon$ and
expanding in orders of $\epsilon$ yields the $\beta$-functions
\be
\beta_g = \epsilon g - b g^2 ,
\label{eq:betag2d}
\ee
with $b=\frac{2}{3}\Big[2 \Phi_0^1(0)+24\Phi_1^2(0)
-\Phi_0^1\big(-\frac{4}{\epsilon}\lambda\big)\Big]$, and
\be
\beta_\lambda = -2\lambda + 2 g\Big[-2\Phi_1^1(0)
+\Phi_1^1\big(-{\textstyle\frac{4}{\epsilon}}\lambda\big)\Big],
\label{eq:betalambda2d}
\ee
where we have dropped higher orders in $\lambda$, $g$ and $\epsilon$, since it is possible to prove
that the fixed point values of both $g$ and $\lambda$ must be of order $\mO(\epsilon)$. Some of the
threshold functions $\Phi$ (cf.\ \cite{R98}) appearing in \eqref{eq:betag2d} and
\eqref{eq:betalambda2d} are independent of the underlying cutoff shape function $R^{(0)}(z)$.
Here we have $\Phi_0^1(0)=1$ and $\Phi_1^2(0)=1$ for any cutoff. Furthermore, for all standard
shape functions satisfying $R^{(0)}(z=0)=1$ we find
$\Phi_0^1\big(-\frac{4}{\epsilon}\lambda\big)=\big(1-\frac{4}{\epsilon}\lambda\big)^{-1}$.
Due to the occurrence of $\epsilon^{-1}$ in the argument of $\Phi_0^1$, the $\lambda$-dependence
does not drop out of $\beta_g$ at lowest order, but rather the combination $\lambda/\epsilon$
results in a finite correction. As an exception, the sharp cutoff \cite{RS02} implicates
$R^{(0)}(z=0)\rightarrow\infty$, leading to $\Phi_0^1\big(-\frac{4}{\epsilon}\lambda\big)
=1$.\footnote{For the sharp cutoff, $\Phi^1_n(w)=-\frac{1}{\Gamma(n)}\ln(1+w)+C$ is determined
up to a constant, which, for consistency, is chosen such that $\Phi^1_n(w=0)$ agrees with
$\Phi^1_n(0)$ for some other cutoff \cite{RS02}. In the limit $n\rightarrow 0$, however, the
$w$-dependence drops out completely, and $\Phi^1_0(w)^\text{sharp}=\Phi^1_0(0)^\text{other}$.
Since $\Phi^1_0(0)=1$ for any cutoff, we find $\Phi^1_0(w)^\text{sharp}=1 \;\, \forall w$.}
Thus, we find
\be
\textstyle
b=\frac{2}{3}\Big[26 - \big(1-\frac{4}{\epsilon}\lambda\big)^{-1} \Big]
\label{eq:bcoeff}
\ee
for all standard cutoffs, and $b=\frac{2}{3} \cdot 25$ for the sharp cutoff. The threshold
function $\Phi_1^1(w)$ is cutoff dependent, also at $w=0$, so $\beta_\lambda$ is nonuniversal, too.
Hence, both $\lambda_*$ and $g_*$ depend on the cutoff.

In order to calculate the critical central charge $\ccr$, we include minimally coupled scalar
fields $\varphi_i$ ($i=1,\ldots,c$) in our analysis, for instance by adding to $\Gamma_k$ in equation
\eqref{eq:EHtrunc} the action $\frac{1}{2} \sum_{i=1}^c\int\dd x\sbg \, \varphi_i\big(-\bar{D}^2\big)
\varphi_i$. In this case, the coefficient $b$ is changed into
$b=\frac{2}{3}\big[26 - (1-\frac{4}{\epsilon}\lambda )^{-1}-c \big]$ for standard shape functions,
and $b=\frac{2}{3}(25-c)$ for the sharp cutoff. The critical value for $c$, determined by the
zero of $b$ at the NGFP, is computed for different shape functions: the optimized cutoff \cite{L01},
the ``$s$-class exponential cutoff'' \cite{LR02} and the sharp cutoff \cite{RS02}. In addition, we
might set $\lambda=0$ by hand in \eqref{eq:bcoeff} such that the result can be compared to the
perturbative studies \cite{KKN93} where the cosmological constant is not taken into account. If we
do so, we find indeed $\ccr=25$, reproducing the perturbative value given in equation
\eqref{eq:ccritNewPert}. For nonvanishing $\lambda$, however, \emph{the critical central charge is
cutoff dependent}, as can be seen in table \ref{tab:ccrit}. In conclusion, the nice (and expected)
result of a \emph{universal} value $\ccr$ found for the standard parametrization cannot be
transferred to the new parametrization. Nevertheless, we obtain a number close to $25$ for all
cutoffs considered, making contact to the CFT result.

{\renewcommand{\arraystretch}{1.2}
\begin{table}[tp]
\begin{tabular}{cc}
 \hline
 Cutoff 						&	$\ccr$			\\
 \hline
 $\quad$Any cutoff, but setting $\lambda=0 \quad$	&	$25$ 			\\
 Optimized cutoff					&	$\quad 25.226 \quad$	\\
 Exponential cutoff ($s=1$)				&	$25.322$		\\
 Exponential cutoff ($s=5$)				&	$25.190$		\\
 Sharp cutoff						&	$25$			\\
 \hline
\end{tabular}
\caption{Cutoff dependence of the critical central charge}
\label{tab:ccrit}
\end{table}%
}%

%-------------------------------------------------------------------------------------------------------

\subsubsection*{Results in \texorpdfstring{$4$}{4} dimensions}

Our analysis in $2+\epsilon$ dimensions suggests that results in the new parametrization might
depend to a larger extent on the cutoff shape. In the following we confirm this conjecture by
considering global properties of the RG flow for different shape functions.

\begin{figure}[htp]
  \centering
  \includegraphics[width=.87\columnwidth]{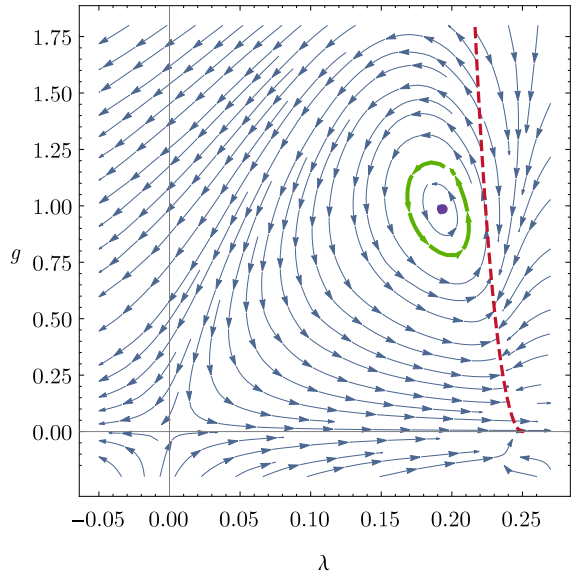}
  \caption{Flow diagram in $d=4$ based on the \emph{new
	  parametrization} and the optimized cutoff.}
  \label{fig:NewSingleOpt}
\end{figure}

\textbf{i)} \emph{Optimized cutoff.} An evaluation of the $\beta$-functions in $d=4$ gives rise
to the flow diagram shown in figure \ref{fig:NewSingleOpt}.

The result is fundamentally different from what is known for the standard parametrization (cf.\
figure \ref{fig:StdSingle}). Although we find again a Gaussian and a non-Gaussian fixed point,
we encounter new properties of the latter. The NGFP is \emph{UV-repulsive} in both directions now.
Furthermore, it is surrounded by a closed UV-attractive \emph{limit cycle}. The singularity line
(dashed) (where the $\beta$-functions diverge and beyond which the truncation ansatz is no longer
reliable) prevents the existence of globally defined trajectories emanating from the limit cycle
and passing the classical regime, i.e.\ there is no connection between the limit cycle and the
Gaussian fixed point. Trajectories inside the limit cycle are asymptotically safe in a generalized
sense since they approach the cycle in the UV, and they hit the NGFP in the infrared, but they can
never reach a classical region. Note that the limit cycle is similar to those found in references
\cite{HR12,DP13} which are based on nonlinear metric parametrizations, too.

\begin{figure}[htp]
  \centering
  \includegraphics[width=.87\columnwidth]{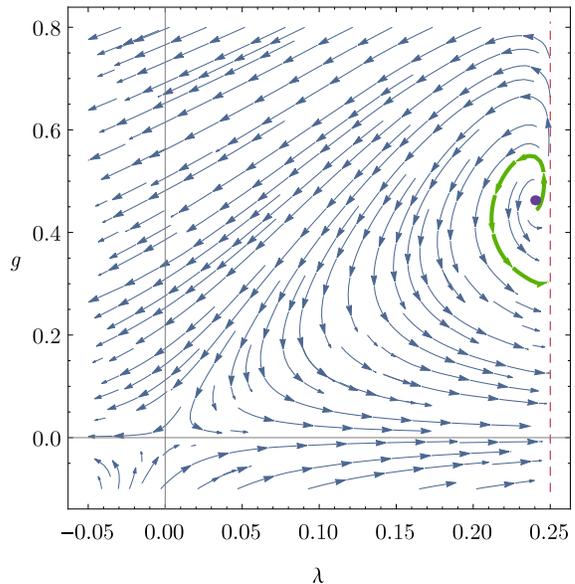}
  \caption{Flow diagram in $d=4$ based on the \emph{new
	  parametrization} and the sharp cutoff.}
  \label{fig:NewSingleSharp}
\end{figure}

\textbf{ii)} \emph{Sharp cutoff.} The flow diagram based on the sharp cutoff -- see figure
\ref{fig:NewSingleSharp} -- is similar to the ones found with the standard parametrization.
There is a Gaussian and a non-Gaussian fixed point. The NGFP is \emph{UV-attractive} in both $g$-
and $\lambda$-direction. There is \emph{no limit cycle}. Due to the singularity line, there is no
asymptotically safe trajectory that has a sufficiently extended classical regime close to the
Gaussian fixed.

\textbf{iii)} \emph{Exponential cutoff.} The exponential cutoff with generic values for the
parameter $s$ gives rise to a flow diagram (not depicted here) that is somewhere in between
figure \ref{fig:NewSingleOpt} and figure \ref{fig:NewSingleSharp}. The NGFP is \emph{UV-repulsive}
as it is for the optimized cutoff. However, there is no closed limit cycle. Although a relict of
the cycle is still present, it does not form a closed line, but rather runs into the singularity
line. Again, there is no separatrix connecting the fixed points. Varying $s$ shifts the
coordinates of the NGFP. For $s\leq 0.93$ \emph{the fixed point even vanishes}, or, more precisely,
it is shifted beyond the singularity, leaving it inaccessible. Thus, the NGFP that seemed to be
indestructible for the standard parametrization can be made disappear with the new parametrization!

In summary, fundamental qualitative features of the RG flow like the signs of critical exponents,
the existence of limit cycles, or the existence of suitable non-Gaussian fixed points are strongly
cutoff dependent.

%-------------------------------------------------------------------------------------------------------

\section{Results: Bi-metric}
\label{sec:bi}

For the bi-metric analysis \cite{bimetric,BR14} we consider the truncation ansatz
\be
\begin{split}
 \Gamma_k\big[g,\bg,\xi,\bxi\, \big] =\; &\frac{1}{16\pi G_k^\text{Dyn}} \int\! \dd x \sg
 \big(\! -R + 2\Lambda_k^\text{Dyn} \big)\\
 &+\frac{1}{16\pi G_k^\text{B}} \int\! \dd x \sbg \big(\! -\bar{R} + 2\Lambda_k^\text{B}
 \big)\\[0.1em]
 &+ \Gamma_k^\text{gf}\big[g,\bg \big] + \Gamma_k^\text{gh}\big[g,\bg,\xi,\bxi\, \big].
\end{split}
\label{eq:doubleEHtrunc}
\ee
It consists of two separate Einstein--Hilbert terms belonging to the dynamical ('Dyn') and the
background ('B') metric with their corresponding couplings. To evaluate $\beta$-functions we
employ the conformal projection technique together with the $\Omega$ deformed $\alpha=1$ gauge
\cite{BR14}.

%-------------------------------------------------------------------------------------------------------

\subsection{Known results for the standard parametrization}
\label{sec:biKnown}

We quote the most important results obtained for the standard parametrization \cite{BR14}.
Since the background couplings $G_k^\text{B}$ and $\Lambda_k^\text{B}$ in the truncation ansatz
\eqref{eq:doubleEHtrunc} occur in terms containing the background metric only, they drop out when
calculating the second derivative of $\Gamma_k$ with respect to $h_\mn$, and hence, they cannot
enter the RHS of the FRGE \eqref{eq:FRGE}. As a consequence, the RG flow of the dynamical coupling
sector is decoupled: $\beta_\lambda^\text{Dyn}\equiv \beta_\lambda^\text{Dyn}(\lambda^\text{Dyn},
g^\text{Dyn})$ and $\beta_g^\text{Dyn}\equiv \beta_g^\text{Dyn}(\lambda^\text{Dyn},g^\text{Dyn})$
form are closed system, so one can solve the RG equations of the 'Dyn' couplings independently
at first.

\begin{figure}[htp]
  \centering
  \includegraphics[width=\columnwidth]{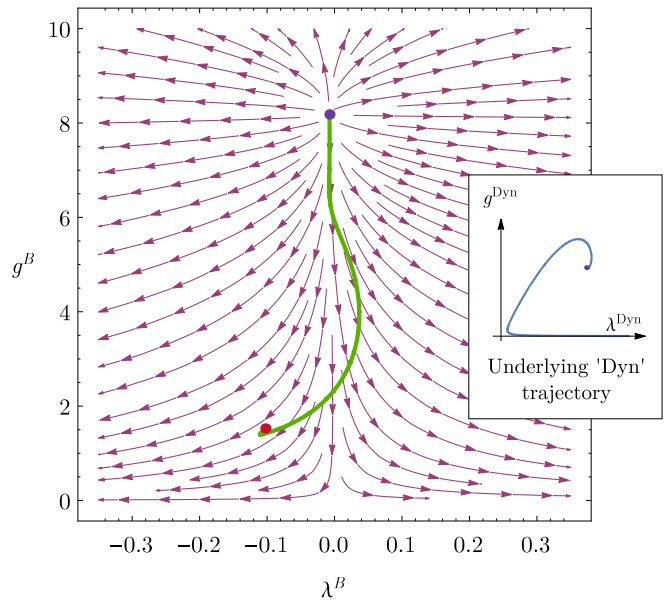}
  \caption{Vector field for the background couplings at $k\rightarrow\infty$ and RG trajectory
	that is asymptotically safe in the UV and restores split symmetry in the IR (left figure),
	and underlying trajectory in the 'Dyn' sector (right figure), based on the \emph{standard
	parametrization} and the optimized cutoff in $d=4$.}
  \label{fig:StdBiOpt}
\end{figure}

On the other hand, the background $\beta$-functions depend on both dynamical and background
couplings. Therefore, the RG running of $g_k^\text{B}$ and $\lambda_k^\text{B}$ can be determined
only if a solution of the 'Dyn' sector is picked. With regard to the Asymptotic Safety program
we choose a 'Dyn' trajectory which emanates from the NGFP and passes the classical regime near the
Gaussian fixed point. This trajectory is inserted into the $\beta$-functions of the background
sector, making them explicitly $k$-dependent. The vector field these $\beta$-functions give rise to
depends on $k$, too, and possible ``fixed points'', i.e.\ simultaneous zeros of
$\beta_\lambda^\text{B}$ and $\beta_g^\text{B}$, become moving points. The UV-attractive ``moving
NGFP'' is called running attractor in \cite{BR14}. In figure \ref{fig:StdBiOpt} we show the vector
field of the background sector at $k\rightarrow\infty$ and the RG trajectory (thick) that starts
at the $k=0$ position of the running attractor and ends at its $k\rightarrow\infty$ position
(w.r.t.\ the inverse RG flow).
As argued in \cite{BR14}, this specific trajectory is the only acceptable possibility once a 'Dyn'
trajectory is chosen: It combines the requirements of Asymptotic Safety (it approaches an NGFP in
the UV) and \emph{split symmetry restoration} in the infrared.

Recovering split symmetry at $k=0$ is of great importance. We mentioned already in the introduction
that the background metric is an auxiliary construction, and physical observables must not depend
on it. Thus, physical quantities derived from the full quantum action $\Gamma=\Gamma_{k=0}$ are
required not to have an extra $\bg$-dependence, but to depend only on $g$. In the standard
parametrization, where $g=\bg+h$, this means that $\bg$ and $h$ can make their appearance only
via their sum (which is split symmetric). Hence, the claim of split symmetry originates from
requiring \emph{background independence}. Within the approximations of \cite{BR14}, this requires
at the level of the effective average action: $1/G_k^\text{B}\rightarrow 0$ and
$\Lambda_k^\text{B}/G_k^\text{B}\rightarrow 0$. For any appropriate choice of initial conditions in
the 'Dyn' sector \emph{there exists a unique trajectory} in the 'B' sector complying with this
requirement in the infrared. This general result is independent of the chosen cutoff shape function.

%-------------------------------------------------------------------------------------------------------

\subsection{Results for the new parametrization}

We aim at finding an asymptotically safe trajectory that restores background independence at
$k=0$. (In the context of the new parametrization we do no longer call it ``split symmetry'' in order
to keep the nonlinearity in mind.) We present some intermediate results of the calculation in appendix
\ref{app:bi}, where we mention the differences to the standard parametrization on the technical level.
In the following, we discuss the differences of the resulting RG flow and its dependence on the cutoff
shape.

\begin{figure}[tp]
  \centering
  \includegraphics[width=.8\columnwidth]{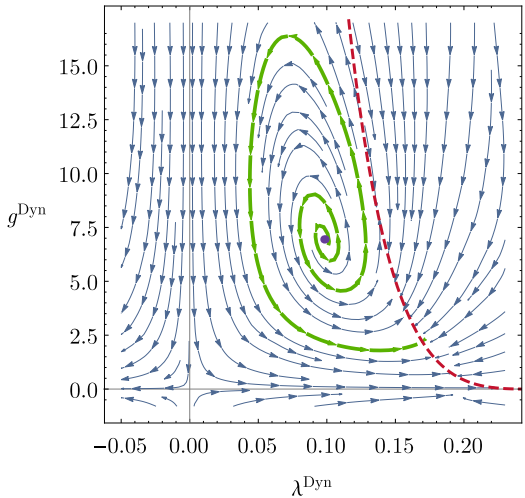}
  \caption{Flow diagram of the 'Dyn' couplings in $d=4$ based on the \emph{new
	  parametrization} and the optimized cutoff.}
  \label{fig:NewBiOptDyn}
\end{figure}

\textbf{i)} \textit{Optimized cutoff.} An evaluation of the $\beta$-functions in the 'Dyn' sector
gives rise to the flow diagram displayed in figure \ref{fig:NewBiOptDyn}.
We discover a non-Gaussian fixed point, but it is rather close to the singularity line. As a
consequence, all trajectories emanating from this fixed point will hit the singularity after a
short RG time. It is \emph{impossible to find suitably extended trajectories}: they do not pass
the classical regime, and they never come close to an acceptable infrared limit. For this reason
it is pointless to investigate the possibility of restoration of background independence.
The background sector exhibits a UV-attractive NGFP, too, but due to the lack of an appropriate
infrared regime we do not show a vector field for the background couplings here.

\textbf{ii)} \textit{Exponential cutoff.} We find the same qualitative picture as in figure
\ref{fig:NewBiOptDyn} which was based on the optimized cutoff. The exponential cutoff brings about
a UV-attractive non-Gaussian fixed point, but there are no trajectories that extend to a suitable
infrared region. Thus, there is \emph{no restoration of background independence}.

\begin{figure}[htp]
	\centering
  \includegraphics[width=\columnwidth]{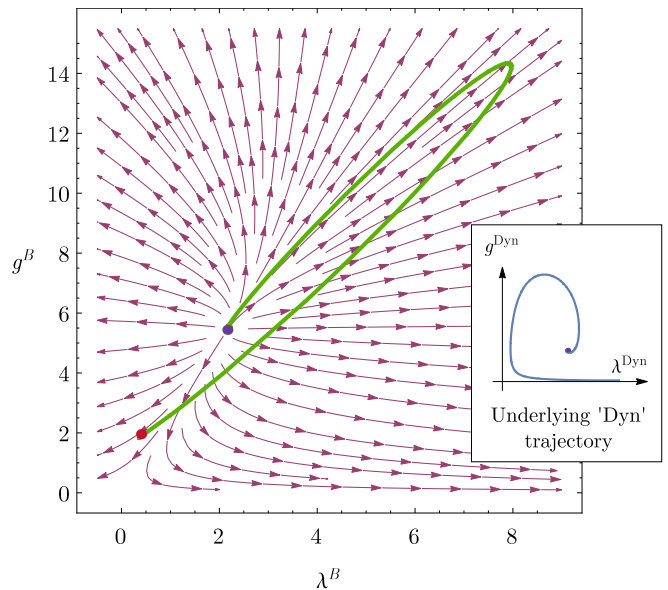}
  \caption{Vector field for the background couplings at $k\rightarrow\infty$ and RG trajectory
	that is asymptotically safe in the UV and restores background independence in the IR (left
	figure), and underlying trajectory in the 'Dyn' sector (right figure), based on the
	\emph{new parametrization} and the sharp cutoff in $d=4$.}
  \label{fig:NewBiSharp}
\end{figure}

\textbf{iii)} \textit{Sharp cutoff.} The $\beta$-functions of the 'Dyn' coup\-lings lead to a
Gaussian and a non-Gaussian fixed point. We observe that $\beta_\lambda^\text{Dyn}$ is
proportional to $\lambda^\text{Dyn}$, so trajectories cannot cross the line at
$\lambda^\text{Dyn}=0$. However, there are trajectories that connect the NGFP to the classical
regime, comparable with the ones found for the standard parametrization. Once such a trajectory
is chosen, it serves as a basis for further analyses since it can be inserted into the
$\beta$-functions of the background sector to study the corresponding RG flow. In this way, we
find the same running attractor mechanism as in section \ref{sec:biKnown} for the standard
parametrization. As can be seen in figure \ref{fig:NewBiSharp}, there is an NGFP present at
$k\rightarrow\infty$, suitable for the Asymptotic Safety scenario.
The thick trajectory is again the unique one that starts at the $k=0$ position of the running
attractor and ends at its $k\rightarrow\infty$ position. Even if the curve has a different form
compared with figure \ref{fig:StdBiOpt}, it has the same essential properties. In particular, it
\emph{restores background independence in the infrared}. Since, in addition, it is asymptotically
safe, it is an eligible candidate for defining a fundamental theory.

To summarize, the possibility of finding a suitable RG trajectory combining the requirements of
Asymptotic Safety and restoration of background independence at $k=0$ depends in a crucial way
on the cutoff shape for the new parametrization.

%-------------------------------------------------------------------------------------------------------

\section{Conclusion}

We have investigated the properties of the ``new'' exponential metric parametrization
$g_{\mu\nu}=\bg_{\mu\rho}(e^h)^\rho{}_\nu$, which has been contrasted with the standard background
field split, $g_{\mu\nu}=\bg_\mn+h_\mn$. When inserting the exponential relation into the classical
Einstein--Hilbert action and expanding in orders of $h_\mn$ we obtain
\be
\begin{split}
S^\text{EH}[g] &= S^\text{EH}\big[\bg e^h\big]=S^\text{EH}\big[\bg+h+\mO(h^2)\big] \\
&= S^\text{EH}[\bg] + \int\dd x\frac{\delta S^\text{EH}}{\delta g_\mn(x)}h_\mn(x) + \mO(h^2).
\end{split}
\ee
Thus, the equations of motion are given by the standard ones,
$\frac{\delta S^\text{EH}}{\delta h_\mn}\big|_{g=\bg}
= \frac{\delta S^\text{EH}}{\delta g_\mn}\big|_{g=\bg}
= \frac{1}{16\pi G}\big(\bar{G}^\mn+\bg^\mn\Lambda\big) = 0$, i.e.\ the two parametrizations
give rise to equivalent theories at the classical level. Since the quantum character of
gravity is not known, there is no reason, a priori, to prefer one parametrization
to another. So far, almost all Asymptotic Safety related studies considered the standard
parametrization. In this work we focused on the new one instead.

We have shown that there is a one-to-one correspondence between dynamical metrics
$g_{\mu\nu}=\bg_{\mu\rho}(e^h)^\rho{}_\nu$ and symmetric fields $h_\mn$ in the exponent, which
is particularly important from a path integral perspective. It remains an open question, however,
if this correspondence can be transferred to Lorentzian spacetimes in the sense that $g_\mn$ and
$\bg_\mn$ have the same signature when related by the new parametrization.

Do we expect different results for the two parametrizations at all? According to the
equivalence theorem, a field redefinition in the path integral does not change S-matrix
elements, but this is an on shell argument, where internal quantum fluctuations are integrated
out completely. In contrast, in the effective average action $\Gamma_k$ fluctuations with
momenta below $k$ are suppressed, and nowhere in the FRGE do we go on shell. Therefore, we
expect differences in $\beta$-functions and structures of the corresponding RG flow, indeed.
Due to the lack of directly measurable physical observables exhibiting quantum gravity effects,
these off shell quantities are of considerable interest.

Clearly, even the role of Newton's constant is changed for the new parametrization.
This can be understood as follows. In order to identify Newton's constant with the strength
of the gravitational interaction in the standard parametrization, one usually rescales
the fluctuations $h_\mn$ such that
\be
g_\mn = \bg_\mn + \sqrt{32\pi G_k}\, h_\mn .
\ee
In this way, each gravitational vertex with $n$ legs is associated with a factor
$(\sqrt{32\pi G_k}\,)^{n-2}$. For the new parametrization we can consider a similar rescaling
of $h_\mn$, leading to the same factor appearing in the $n$-point functions. The difference
resides in the fact that there are new terms and structures in $\Gamma_k^{(n)}$ when using
the new parametrization. As already indicated in equation \eqref{eq:2ndVar}, these
additional contributions to each vertex are due to the chain rule. Hence, Newton's constant
is accompanied by different terms in the $n$-point functions.

In fact, these general considerations are reflected in our findings. To summarize them: 1.) We
find quite different $\beta$-functions and new structures in the RG flow. 2.) The calculations
based on a type I cutoff result in a strong dependence on the cutoff shape function.

This can be seen most clearly in $d=4$ dimensions. In the single-metric computation we encountered
a limit cycle and a UV-repulsive NGFP for the optimized cutoff, whereas the sharp cutoff gives
rise to a UV-attractive NGFP without limit cycle. Furthermore, in the bi-metric setting with a
sharp cutoff there exists an asymptotically safe trajectory that restores background independence
in the infrared, while it is not possible to find such trajectories when using the optimized
cutoff. It is remarkable and somewhat unexpected that the sharp cutoff leads to the most convincing
results.

Our observations seem to suggest that results based on the new parametrization are less
reliable or even unphysical. On the other hand, the strong cutoff dependence compared to the
linear parametrization could be seen from a different perspective as well: If Quantum Einstein
Gravity with $g_{\mu\nu}=\bg_{\mu\rho}(e^h)^\rho{}_\nu$ is asymptotically safe, probably more
invariants in the truncation ansatz are needed to get a clear picture. The nonlinear relation
for the metric might give more importance to the truncated higher order terms.

Moreover, it can be speculated that the strong dependence on the cutoff shape might be a
peculiarity of the type I cutoff. As has been argued in \cite{DP13}, in some situations the type
II cutoff leads to correct physical results, whereas the type I cutoff does not. Future calculations
may show if a similar reasoning applies here as well, i.e.\ if the essential properties of the RG
flow obtained with a type II cutoff do not depend on the shape function to such a great extent.
We conjecture that the limit cycle is a consequence of a nonlinear parametrization in
combination with a type I cutoff. Instead, a type II cutoff might lead to physical and stable
results so that the advantages of the new parametrization become more apparent.

In $d=2+\epsilon$ dimensions we can reproduce the critical value of the central charge
obtained with a perturbative calculation, $\ccr=25$, when the cosmological constant $\lambda$ is
set to zero. For nonvanishing $\lambda$ we find a slight cutoff dependence of $\ccr$, but it
remains still close to $25$. Since it is this number that makes contact to vertex operator
calculations and other established CFT arguments, the new parametrization seems to be appropriate
for comparisons and further applications after all.

%-------------------------------------------------------------------------------------------------------

\begin{acknowledgments}
The author would like to thank M.~Reuter and S.~Lippoldt for many extremely useful suggestions.
He is also grateful to H.~Gies, R.~Percacci, A.~Codello, K.~Falls, D.~Benedetti and A.~Eichhorn
for stimulating discussions at the ERG conference 2014. 
\end{acknowledgments}

%-------------------------------------------------------------------------------------------------------

\appendix

\section[One-to-one correspondence between $g$ and $h$]
{One-to-one correspondence between \texorpdfstring{\bm{$\lowercase{g}$}}{g}
and \texorpdfstring{\bm{$\lowercase{h}$}}{h}}
\label{app:Logs}

In this appendix we show that there is a one-to-one correspondence between (Riemannian)
dynamical metrics $g_\mn$ and symmetric matrix-valued fields $h_\mn$ whose relation is given
by $g_\mn=\bg_{\mu\rho}(e^h)^\rho{}_\nu$. In order to transform this from component notation
to matrix notation we note that the $h$ appearing in the exponent has implicit index
positions $h^\rho{}_\nu$, signaling the fact that it is actually a product of $h$ and the
inverse background metric, $h^\rho{}_\nu=\bg^{\rho\sigma}h_{\sigma\nu}$, so we have
$h^\rho{}_\nu=(\bg^{-1}h)^\rho{}_\nu$. Thus, the full relation in matrix form reads
\begin{equation}
g=\bg \, e^{\bg^{-1}h} \, .
\label{eq:paramMatrix}
\end{equation}
Note that statements about symmetry and positive definiteness are not trivial since the
product of symmetric positive definite matrices is in general neither positive definite
nor symmetric. A priori, there is also not much known about real logarithms of products
of matrices. Here we need the following theorem concerning existence and uniqueness
of real symmetric matrix logarithms.
\medskip

\noindent
\textbf{Theorem 1.} Let $C$ be a real symmetric positive definite matrix. Then there exists
a unique real symmetric solution $H$ to the equation $C=e^H$.
\medskip

\noindent
\textbf{Proof}\\
\textit{Existence}: Since $C\in\text{Sym}_{n\times n\,}$, there exist an $S\in\On$ and a
diagonal matrix $\Lambda=\diag(\lambda_1,\ldots,\lambda_n)$, with $\{\lambda_i\}$ the
eigenvalues of $C$, such that $C=S^T\Lambda S$. Positive definiteness of $C$ implies that
all $\lambda_i$ be positive. Now, set $H=S^T \diag(\ln\lambda_1,\ldots,\ln\lambda_n) S$.
Then $H$ is real and symmetric. Exponentiating $H$ yields
\begin{align*}
	e^H = S^T e^{\diag(\ln\lambda_1,...,\ln\lambda_n)}S
	= S^T\diag(\lambda_1,...,\lambda_n)S =C,
\end{align*}
	proving the existence of a real symmetric solution.\\
\textit{Uniqueness}: Assume $H$ is a real symmetric matrix satisfying $C=e^H$. Assume $H'$ is another
real symmetric matrix with the same exponential, $C=e^{H'}$. Due to the symmetry, there are matrices
$O\in\On$ and $O'\in\On$ together with the diagonal matrices $D=\diag(d_1,\ldots,d_n)$ and
$D'=\diag(d_1',\ldots,d_n')$, where $d_i$ ($d_i'$) are the eigenvalues of $H$ ($H'$), such that
$H=O^TDO$ and $H'={O'}^TD'O'$. Then we have $C=e^H=e^{O^TDO}=O^Te^DO$, and, similarly,
$C={O'}^Te^{D'}O'$. Equating these expression leads to $e^D\big(O{O'}^T\big)=\big(O{O'}^T\big)e^{D'}$,
or, rewritten,
\begin{equation}
 e^D U=U e^{D'} ,
\label{eq:UD}
\end{equation}
with $U=O{O'}^T\in\text{O}(n)$. The matrix entries in \eqref{eq:UD} are
\begin{equation}
 \big( e^{D} U \big)_{ij} = \sum\limits_{k=1}^{n} e^{d_{i}} \delta_{i k} u_{k j}
 = e^{d_{i}} u_{i j} \, ,
\end{equation}
and, analogously, $\big( U e^{D'} \big)_{ij} = e^{d'_{j}} u_{i j}$.
For any pair $(i,j)$ this gives the relation $(e^{d_{i}} - e^{d'_{j}} ) u_{i j} = 0$.
Since all $d_{i}$ are real, we conclude that $( d_{i} - d'_{j} ) u_{i j} = 0$. Again in matrix
form this reads $D U -  U D' = 0$. Re\-instating $U=O{O'}^T$ and rearranging finally results in
\begin{equation}
H=O^TDO={O'}^TD'O'=H' \, ,
\end{equation}
which proves the uniqueness of $H$.
\medskip

Now let us come back to equation \eqref{eq:paramMatrix}. First we consider the case where
(apart from $\bg$) $h$ is given.
\medskip

\noindent
\textbf{Theorem 2.} If $h$ is real and symmetric, then $g$ defined by
$g=\bg\, e^{\bg^{-1}h}$ is symmetric and positive definite.
\medskip

\noindent
\textbf{Proof}\\
\textit{Symmetry}:
\begin{align}
 g^T&=\big(e^{\bg^{-1}h}\big)^T\bg^T=e^{h^T(\bg^{-1})^T}\bg=e^{\bg\,\bg^{-1}h\,\bg^{-1}}\,\bg
 \nonumber\\
 &=\bg\, e^{\bg^{-1}h}\,\bg^{-1}\bg=\bg\, e^{\bg^{-1}h}=g \, .
\end{align}
\textit{Positive definiteness}: Since $\bg$ is symmetric and positive definite, we can define
$\bg^{1/2}$ to be the (unique) principal square root, which is again symmetric. Let $z$ be a nonzero
vector with $n$ real components. Then
\begin{equation}
\begin{split}
z^Tgz&=z^T \bg\, e^{\bg^{-1}h} z = z^T \bg\, e^{\bg^{-1/2}\bg^{-1/2}h\,\bg^{-1/2}\bg^{1/2}} z\\
&= z^T \bg\,\bg^{-1/2} e^{\bg^{-1/2}h\,\bg^{-1/2}}\bg^{1/2} z\\
&= (\bg^{1/2}z)^T e^{\bg^{-1/2}h\,\bg^{-1/2}}(\bg^{1/2} z) = y^T e^Ky \, ,
\end{split}
\end{equation}
with $y=\bg^{1/2} z$ and $K=\bg^{-1/2}h\,\bg^{-1/2}$. We observe that $K$ is symmetric.
Thus, we can write $e^K=\big(e^{\frac{1}{2}K}\big)^Te^{\frac{1}{2}K}$.
Inserting this into the previous equation yields
\begin{equation}
z^Tgz = y^T e^Ky = \big(e^{\frac{1}{2}K}y\big)^T \big(e^{\frac{1}{2}K}y\big) = x^Tx>0 \,,
\end{equation}
with $x=e^{\frac{1}{2}K}y$, proving positive definiteness.
\medskip

Finally we consider the case where $g$ and $\bg$ are given.
\medskip

\noindent
\textbf{Theorem 3.} Let $\bg$ be the background metric and $g$ be any dynamical metric at
a given spacetime point $x$. Then there exists a unique real symmetric matrix $h$ satisfying
$g=\bg\, e^{\bg^{-1}h}$.
\medskip

\noindent
\textbf{Proof}\\
\textit{Existence}: Let again $\bg^{1/2}$ be the (real and symmetric) principal square root of
$\bg$. The key idea is to rewrite the defining equation as follows.
\begin{equation}
\begin{split}
 g&=\bg\, e^{\bg^{-1}h} = \bg\, e^{\bg^{-1/2}\bg^{-1/2}h\,\bg^{-1/2}\bg^{1/2}}\\
	&= \bg^{1/2} e^{\bg^{-1/2}h\,\bg^{-1/2}}\bg^{1/2},
\end{split}
\end{equation}
leading to
\begin{equation}
\bg^{-1/2} g\, \bg^{-1/2} = e^{\bg^{-1/2}h\,\bg^{-1/2}}.
\label{eq:paramMatrixModified}
\end{equation}
We observe that the left hand side of equation \eqref{eq:paramMatrixModified} is symmetric and
positive definite ($z^T\bg^{-1/2} g\, \bg^{-1/2}z=(\bg^{-1/2}z)^T g\, (\bg^{-1/2}z)=y^Tgy>0$).
According to theorem 1, there exists a unique real symmetric matrix $H$
satisfying $\bg^{-1/2} g\, \bg^{-1/2}=e^H$. Setting $h=\bg^{1/2}H\,\bg^{1/2}$
and noting that $h$ is real and symmetric proves the existence.

\noindent
\textit{Uniqueness}: Since there is more than one square root of $\bg$ in general,
it remains to show that the $h$ constructed above does not depend on the choice
of the root. Assume there exists another symmetric solution $h'$ corresponding to
another square root $(\bg^{1/2})'$, i.e.\ $g = \bar{g}\, e^{\bar{g}^{-1} h'}$.
In the manner of equation \eqref{eq:paramMatrixModified} we rewrite again
\begin{equation}
 \bg^{-1/2} g\, \bg^{-1/2} = e^{\bg^{-1/2} h'\, \bg^{-1/2}}
\stackrel{!}{=} e^{\bg^{-1/2} h\, \bg^{-1/2}},
\end{equation}
where we use the principal root $\bg^{1/2}$ on all sides. We already know that the symmetric
logarithm of the LHS is unique. Therefore, the exponents on the RHS have to agree,
$\bg^{-1/2} h'\, \bg^{-1/2} = \bg^{-1/2} h\, \bg^{-1/2}$, and
finally $h'=h$, completing the proof of uniqueness.

%-------------------------------------------------------------------------------------------------------

\section[Hessians and $\beta$-functions]{Hessians and \texorpdfstring{\bm{$\beta$}}{beta}-functions}
\label{app:Details}

\subsection{Single-metric}
\label{app:single}

We adopt the notation of reference \cite{R98}. When inserting the new parametrization
\eqref{eq:newParam} into $\Gamma_k$ given by equation \eqref{eq:EHtrunc} and expanding $\Gamma_k$ in
orders of $h_\mn$, the quadratic term reads
\be
 \Gamma_k^\text{quad} = \frac{1}{32 \pi G_k} \int \dd x \sbg \, h_\mn \! \left( 
 -K^\mn{}_\rs \bar{D}^2 + U^\mn{}_\rs \right) h^\rs,
\ee
with $K^\mn{}_\rs=\frac{1}{4}(\delta^\mu_\rho \delta^\nu_\sigma +
\delta^\mu_\sigma \delta^\nu_\rho - \bg^\mn \bg_{\rho\sigma})$ and
\be
\begin{split}
U^\mn{}_\rs= & -\frac{1}{4} \, \bg^\mn \bg_\rs \bar{R} + \frac{1}{2}\big(\bg^\mn \bar{R}_\rs
+ \bg_\rs \bar{R}^\mn\big)\\
& - \frac{1}{2}\big(\bar{R}^\nu{}_\rho{}^\mu{}_\sigma + \bar{R}^\nu{}_\sigma{}^\mu{}_\rho\big)
+ \frac{1}{2} \, \bg^\mn \bg_\rs \Lambda_k \, ,
\end{split}
\ee
so the additional terms resulting from the new para\-metrization cancel some of those which are
already present in the standard calculation. After splitting the field $h_\mn$ into trace and
traceless part, $h_\mn=\hat{h}_\mn+\frac{1}{d}\bg_\mn \phi$, where $\phi=\bg^\mn h_\mn$ and
$\bg^\mn\hat{h}_\mn=0$, and inserting a maximally symmetric background for $\bg_\mn$, we obtain
\be
\begin{split}
\Gamma_k^\text{quad} = \frac{1}{64\pi G_k} \int\dd x\sbg \, \bigg\{ \hat{h}_\mn \Big( -\bar{D}^2
+ C_\text{T} \bar{R} \Big) \hat{h}^\mn \\
- \bigg(\frac{d-2}{2d}\bigg) \phi \Big(-\bar{D}^2 + C_\text{S} \bar{R} - \mu \Lambda_k \Big)\phi
\bigg\},
\end{split}
\ee
with the constants $C_\text{T}=\frac{2}{d(d-1)}$, $C_\text{S}=\frac{d-2}{d}$ and
$\mu=\frac{2d}{d-2}$. As argued on general grounds in the introduction, the cosmological
constant does indeed drop out of the trace\-less sector.

The resulting anomalous dimension of Newton's constant is given by
$\eta_N=\frac{g B_1}{1-g B_2}$, where
\begin{align}
B_1 =& \quad\frac{1}{3}(4\pi)^{-\frac{d}{2}+1} \bigg\{ \big(d^2-3d-2\big) \Phi_{d/2-1}^1(0) 
\nonumber\\
&\mkern6mu -12\,\frac{3d+2}{d}\, \Phi_{d/2}^2(0) + 2\,\Phi_{d/2-1}^1(-\mu\lambda) \\
&\mkern108mu -12\,\frac{d-2}{d}\,\Phi_{d/2}^2(-\mu\lambda) \bigg\} , 
\nonumber \displaybreak[0]\\
B_2 =& -\frac{1}{6}(4\pi)^{-\frac{d}{2}+1} \bigg\{ (d-1)(d+2) \tilde{\Phi}_{d/2-1}^1(0)
\nonumber\\
&\mkern23mu -12\,\frac{d+2}{d}\, \tilde{\Phi}_{d/2}^2(0)
+ 2\,\tilde{\Phi}_{d/2-1}^1(-\mu\lambda) \\
&\mkern116mu -12\,\frac{d-2}{d}\,\tilde{\Phi}_{d/2}^2(-\mu\lambda) \bigg\}. \nonumber
\end{align}
The threshold functions $\Phi$ and $\tilde{\Phi}$ are defined in reference \cite{R98}.
Finally, we have the $\beta$-functions of $g_k=k^{d-2}G_k$ and $\lambda_k=k^{-2}\Lambda_k$,
\begin{align}
 \beta_g = &(d-2+\eta_N)g,
 \label{eq:beta_g_FRG}\\
 \beta_\lambda = &-(2-\eta_N)\lambda \nonumber\\
 & + {\textstyle \frac{1}{2}}(4\pi)^{-\frac{d}{2}+1} g \, \Big \{
 2\big(d^2-3d-2\big) \Phi_{d/2}^1(0) \nonumber\\ &- (d-1)(d+2)\eta_N \tilde{\Phi}_{d/2}^1(0)
 \nonumber\\ &+ 4\Phi_{d/2}^1(-\mu\lambda) - 2\eta_N \tilde{\Phi}_{d/2}^1(-\mu\lambda) \Big\}.
 \label{eq:beta_lambda_FRG}
\end{align}

%-------------------------------------------------------------------------------------------------------

\subsection{Bi-metric}
\label{app:bi}

The conformal projection technique consists in setting the dynamical metric to
$g_\mn=e^{2\Omega}\bg_\mn$. In the following, we denote this projection by $(\cdots)|_\text{pr}\,$.
For the new parametrization, $g_{\mu\nu}=\bg_{\mu\rho}(e^h)^\rho{}_\nu$, it is equivalent to setting
$h^\rho{}_\nu=2\Omega\,\delta^\rho_\nu\,$. This affects the derivatives of $g_\mn$ w.r.t.\
$h_{\rho\sigma}$ appearing in equation \eqref{eq:2ndVar} as follows,
\begin{align}
&\frac{\delta g_\mn(x)}{\delta h_{\rho\sigma}(y)}\, \bigg|_\text{pr}
= e^{2\Omega} \, \delta^\rho_\mu \, \delta^\sigma_\nu \, \delta(x-y), \displaybreak[0]\\[0.2em]
&\frac{\delta^2 g_\mn(u)}{\delta h_{\rho\sigma}(x) \, \delta h_{\lambda\gamma}(y)}
\, \bigg|_\text{pr} \\ 
&= {\textstyle \frac{1}{2}}\, e^{2\Omega}
\left(\bg^{\sigma\lambda} \delta^\rho_\mu \, \delta^\gamma_\nu + \bg^{\rho\gamma} \delta^\lambda_\mu
\, \delta^\sigma_\nu \right) \delta(u-x) \delta(u-y) . \nonumber
\end{align}
After applying the conformal projection and choosing the $\Omega$ deformed $\alpha=1$ gauge
for the same gauge fixing action as in reference \cite{BR14} (or, more precisely, its
``$g_\mn$-version'', cf.\ section \ref{sec:framework}), we obtain the Hessian
\be
\begin{split}
 \big(\Gamma_k^{(2)}\big)^{\mn\rho\sigma}\big|_\text{pr} =
 \frac{e^{(d-2)\Omega}}{32\pi G_k^\text{Dyn}} \Big\{ \big(-\bg^{\mu\rho}\bg^{\nu\sigma}+
 \half\bg^\mn\bg^\rs \big)\bar{D}^2 \\
 -\half\big(\bar{R}-2 e^{2\Omega}\Lambda_k^\text{Dyn}\big) \bg^\mn \bg^\rs \\
 + 2\bar{R}^{\rho\mu\nu\sigma} + \bg^\rs \bar{R}^\mn +\bg^\mn \bar{R}^\rs \Big\}
\end{split}
\label{eq:biHessian}
\ee
in the graviton sector, and
\be
\Big(\big(\Gamma_k^\text{gh}\big)_{\xi\bxi}^{(2)}\Big)^\mu_{~\nu} = \sqrt{2} e^{2\Omega}
\big(\bar{R}^\mu{}_\nu + \delta^\mu_\nu \bar{D}^2\big)
\ee
and $\big(\Gamma_k^\text{gh}\big)_{\bxi\xi}^{(2)}=-\big(\Gamma_k^\text{gh}\big)_{\xi\bxi}^{(2)}$
in the ghost sector. Compared to \cite{BR14}, the Hessians for the ghosts have not changed,
but the one for the graviton is different: (a) The terms in the curly brackets in
\eqref{eq:biHessian} have changed, in particular, the cosmological constant term is
proportional to $\bg^\mn\bg^\rs$ now, so it drops out of the traceless sector as it did in
the single-metric computation of section \ref{app:single}. (b) The numerator of the prefactor
has changed from $e^{(d-6)\Omega}$ into $e^{(d-2)\Omega}$, signaling the special role of $2$
dimensions. We include this factor also in the cutoff $\Rk$.

The $\beta$-functions are determined as in \cite{BR14} by inserting the Hessians into the FRGE
and projecting the trace onto the corresponding invariants. Since they contain a large number of
terms, it would cost several pages to write them down, and we refrain from presenting them here.
Instead, we show the resulting flow diagrams and analyze their properties in section \ref{sec:bi}.

%-------------------------------------------------------------------------------------------------------

\end{document}